\documentclass[preprint,showpacs,preprintnumbers,amsmath,amssymb,nofootinbib]{revtex4}
\pdfoutput=1
\usepackage{booktabs}
\usepackage{mathrsfs}
\usepackage{epsfig}
\usepackage{graphicx}
\usepackage{dcolumn}
\usepackage{bm}
\usepackage{amsmath}
\usepackage{slashed}
\usepackage{subfigure}
\usepackage{ulem}
\usepackage{float}
\usepackage{color}

\usepackage{ulem}

\begin{document}

\preprint{
\begin{minipage}[b]{1\linewidth}
\begin{flushright}
IPMU16-0179
 \end{flushright}
\end{minipage}
}

\title{Status of CMSSM in light of current LHC Run-2 and LUX data}

\author{Chengcheng Han$^{1}$, Ken-ichi Hikasa$^{2}$,  Lei Wu$^{3,4}$,   Jin Min Yang$^{5,6,2}$,
        Yang Zhang$^{5,6}$
        \\~ \vspace*{-0.3cm} }

\affiliation{
 $^1$ Kavli IPMU (WPI), The University of Tokyo, Kashiwa, Chiba 277-8583, Japan\\
 $^2$ Department of Physics, Tohoku University, Sendai 980-8578, Japan\\
 $^3$ ARC Centre of Excellence for Particle Physics at the Terascale, School of Physics,
      The University of Sydney, NSW 2006, Australia\\
 $^4$ Department of Physics and Institute of Theoretical Physics, Nanjing Normal University,
      Nanjing 210023, China\\
 $^5$ CAS Key Laboratory of Theoretical Physics, Institute of Theoretical Physics,
      Chinese Academy of Sciences, Beijing 100190, China\\
 $^6$ School of Physics, University of Chinese Academy of Sciences, Beijing 100049, China
     \vspace*{1.5cm}}

\begin{abstract}
Motivated by the latest results of the LHC Run-2 and LUX experiments, we examine the status of the 
constrained minimal supersymmetric standard model by performing a global fit. We construct a 
likelihood function including the electroweak precision observables, $B$-physics measurements, LHC Run-1 
and -2 data of SUSY direct searches, Planck observation of the dark matter relic density and the combined 
LUX Run-3 and -4 detection limits. Based on the profile likelihood functions of 1 billion samples, we obtain 
the following observations: 
(i)  The stau coannihilation region has been mostly excluded by the latest LHC Run-2 data; 
(ii) The focus point region has been largely covered by the LUX-2016 limits while the $A$-funnel region  
has been severely restricted by flavor observables like $B_s \to \mu^+\mu^-$. The remaining parts of both 
regions will be totally covered by the future LZ dark matter experiment; 
{(iii) Part of the stop coannihilation region may be detected with higher integrated luminosity LHC;
(iv) The StauC$\&$AF hybrid region still survives considering both LHC and dark matter experiments;}
{(v) The masses of the stop, the lightest neutralino and the gluino have been pushed up to 363 GeV, 
328 GeV and 1818 GeV, respectively.}
\end{abstract}

\maketitle

\section{Introduction}
Supersymmetry (SUSY) is one of the most promising candidates for addressing the long outstanding hierarchy 
problem in particle physics. Among various SUSY models, the constrained minimal supersymmetric standard 
model (CMSSM) \cite{Kane:1993td} is especially attractive, which is an effective parametrization motivated 
by supergravity models. In CMSSM, there are only five free parameters at the GUT scale: the universal 
scalar mass $M_0$, the gaugino mass parameter $M_{1/2}$, the universal trilinear coupling $A_0$, the ratio 
of the two Higgs vacuum expectation values $\tan\beta$ and the sign of the Higgs/higgsino mass parameter 
$\rm{sign}(\mu)$. All the masses and couplings of the sparticles at weak scale are determined by these 
five input parameters through the running of renormalization group equations (RGEs), which makes the 
CMSSM highly predictive.

In the past decades, many efforts have been devoted to investigate the CMSSM from various collider and 
dark matter experiments. In particular, the discovery of the 125 GeV Higgs boson \cite{Aad:2012tfa,Chatrchyan:2012xdj} 
and null results of the LHC direct sparticle searches provide significant constraints on the 
parameter space of the CMSSM \cite{Kowalska:2013hha,Roszkowski:2014wqa,Kowalska:2015kaa,Bagnaschi:2015eha,Bechtle:2015nua,Bechtle:2014yna,Beskidt:2014oea,Kowalska:2014hza,Ellis:2013oxa,Buchmueller:2013rsa,Bechtle:2013mda,Henrot-Versille:2013yma,Bornhauser:2013aya,Citron:2012fg,Desai:2014uha,Strege:2012bt,Buchmueller:2012hv,Cao:2012yn,Fowlie:2012im,Balazs:2013qva,Bechtle:2012zk,Cao:2011sn}. The value of $M_0$ has been pushed up to several hundreds GeV or larger, 
and thus leads to a tension between the observed Higgs mass and muon $g-2$ anomaly. In order to be 
compatible with the measured cold dark matter relic density, the parameter space of the CMSSM is further 
restricted to several regions, namely stau coannihilation (StauC) strip, {stop coannihilation (StopC) strip,} 
focus-point (FP) region where the annihilation rate is 
enhanced by a significant higgsino component in the lightest neutralino, and $A$-funnel (AF) at large 
$\tan\beta$ where the annihilation via a heavy MSSM Higgs boson $H/A$ is dominant. Among these regions, 
the FP region was expected to have a low fine-tuning due to the RGE trajectories of the mass squared 
of the Higgs doublet ($m^2_{H_2}$) crossing close to the electroweak scale \cite{feng-1}. However, the 
large value of $M_0$ required by the Higgs mass indicates a $\sim 0.1\%$ fine tuning in FP 
region \cite{feng-2,feng-3}.

Very recently, the ATLAS and CMS collaborations updated their results of searching for the sparticles 
at the 13 TeV LHC with an integrated luminosity of 13 fb$^{-1}$. The non-excess of multi-jets plus 
missing transverse momentum events excluded the gluino up to 1.8 TeV for a massless neutralino LSP \cite{ATLAS:2016uzr,ATLAS:2016kts}, which is much stronger than the bound of $\sim 1.3$ TeV from the LHC Run-1 \cite{Aad:2014wea}. Meanwhile, the LUX experiments reported the limits on the spin-independent dark matter-nucleon 
scattering using a $3.35\times 10^4$ kg-day exposure \cite{Akerib:2016vxi}. Their new limits are about 
four times stronger than the LUX-2013 results \cite{Akerib:2013tjd}. All of these results will tightly 
constrain the parameter space of the CMSSM and illuminate the way to the search of CMSSM in 
future experiments. Therefore, it is necessary to examine the CMSSM with these new experimental data.

In this work, we adopt a Bayesian approach to scan the parameter space of the CMSSM by constructing a global
likelihood function. We implement the constraints including the Higgs boson mass, the dark matter relic 
density, the flavor observables, the electroweak precision data and the muon $g-2$ measurement. We will 
compare the currently allowed parameter space of the CMSSM with the previous result allowed by the LHC Run-1 
data and present the lower limits for various sparticles. We also estimate the prospects of the LUX-Zeplin 
(LZ) experiment and the high luminosity LHC for covering the parameter space of the CMSSM. 

The paper is organized as follows. In Section II we describe the scan strategy and the constraints used 
in this study. In Section III we present the allowed parameter space of CMSSM under the current constraints 
and show the prospects of future experiments. We draw our conclusion in Section IV.

\section{Scan Strategy and Constraints} 
In our scan, we use the Markov Chain Monte Carlo (MCMC) method based on the Metropolis-Hastings 
algorithm \cite{deAustri:2006pe} to obtain the samples. The MCMC algorithm is used to generate a chain 
of samples whose density is proportional to the posterior probability density function (PDF) $p(\eta|d)$. 
The PDF represents the state of knowledge about the parameter $\eta$ after the experimental data $d$, which 
is given as
\begin{equation}\label{eq:bayes}
\begin{aligned}
  p(\eta|d)=\frac{p(d|\epsilon(\eta))\pi(\eta)}{p(d)},
\end{aligned}
\end{equation}
where $\epsilon(\eta)$ stands for some experimental observable. The likelihood function 
$p(d|\epsilon(\eta))\equiv \mathcal{L}(\eta)$ gives the probability density of obtaining $d$ from 
experiential measurements of $\epsilon$. The prior PDF $\pi(\eta)$ parameterizes
the assumptions about the theory before performing the measurements, and the evidence $p(d)$ represents
the assumptions on the data.

\subsection{Prior PDF}
In the prior PDF $\pi(\eta)$, $\eta$ represents the parameters of CMSSM and SM. For our study, they are 
$M_0$, $M_{1/2}$, $A_0$, tan$\beta$, sign($\mu$) and the top quark mass $m_t$. The prior PDF is chosen 
subjectively to concentrate on the parameter space of interests. There are two popular choices for the prior 
PDF, which are the so-called \textit{Flat} and \textit{Log} scans. The \textit{Flat} prior means that the 
PDF for the parameter is uniform in the given range, while \textit{Log} means that the PDF of parameter is 
logarithm. Thus, \textit{Log} prior has more preference in the lower parameter region. The SM parameter 
$m_t$ is given by the experimental data, and therefore we adopt Gaussian prior, 
$\pi(\eta)=e^{-\frac{(\eta-\hat{\eta})^2}{2\sigma^2}}$. In Table \ref{tab:Ranges}, we display the prior PDF and 
the scan ranges of the CMSSM parameters. Other SM parameters are taken as \cite{PDG}
\begin{eqnarray}
m_b(m_b)=4.18, \quad \alpha_s(m_Z)=0.1185, \quad [\alpha_{EM}(m_Z)]^{-1}=127.944.
\end{eqnarray}

\begin{table}[th] \centering\caption{The parameter space scanned in our MCMC sampling.\label{tab:Ranges}}
\centering
\begin{tabular}{|c|c|c|c|c|c|c|c|c|c|}
\hline Parameter ~&~ $M_0$ (GeV) ~&~ $M_{1/2}$ (GeV) ~&~ $A_0$ (TeV)   ~&~  tan$\beta$    ~&~ sign($\mu$) ~&~ $m_t$ \\
\hline Prior PDF  ~&~  Flat, Log         ~&~ Flat, Log              ~&~ Flat                   ~&~ Flat              ~&~ Fixed           ~&~  Gaussian \\
\hline Range       ~&~  (100,~10000)    ~&~(100,~4000)       ~&~ ~(-10,~10)~ ~&~ (2.0,~65.0)     ~&~  $\pm$1               ~&~  172.9$\pm$0.91\\
\hline
\end{tabular}
\end{table}

\subsection{Likelihood function}

We construct the likelihood function $\mathcal{L}(\eta)$ in Eq.(\ref{eq:bayes}) by using the electroweak 
precision observables, the $B$-physics observables, the measured Higgs boson mass, the Planck observation 
of the cold dark matter relic density ($\mathcal{L}_{\textrm{precision}}$), the null results of SUSY searches 
at the LHC ($\mathcal{L}_{\textrm{LHC}}$) and LEP ($\mathcal{L}_{\textrm{LEP}}$), and the limits of the 
LUX-2016 spin-independent DM scattering cross section ($\mathcal{L}_{\textrm{LUX}}$):
\begin{equation}
    \ln\mathcal{L}(\eta)=\ln\mathcal{L}_{\textrm{precision}}+\ln\mathcal{L}_{\textrm{LHC}}+\ln\mathcal{L}_{\textrm{LEP}}+\ln\mathcal{L}_{\textrm{LUX}}
\end{equation}
In Table \ref{tab:experiment}, we list the experimental values used in calculating $\mathcal{L}_{\textrm{precision}}$, in which we assume the Gaussian approximation:
\begin{equation}
    \ln\mathcal{L}_{\textrm{precision}}=- \mathop{\sum}_{i} \frac{[\hat{\mu}_i(\eta)-\mu_i]^2}{2(\sigma_i^2+\tau_i^2)},
\end{equation}
where $\hat{\mu}_i(\eta)$ denotes the predicted value of observable $\eta$ in the model, $\mu_i$ is the 
corresponding central value of experimental measurement, $\sigma_i$ and $\tau_i$ are experimental and theory 
uncertainties, respectively. In our study, we use \textsf{SOFTSUSY 3.3.1} \cite{Allanach:2001kg} to generate
the mass spectrum of sparticles. We use the \textsf{MicrOMEGAs 2.4.5} \cite{Belanger:2001fz} to calculate 
the dark matter relic density by assuming that the lightest neutralino is the solely dark matter in the 
universe. We also employ \textsf{SuperIso v3.3} \cite{Mahmoudi:2008tp} to evaluate the flavor physics 
observables and \textsf{FeynHiggs 2.12} \cite{Heinemeyer:1998yj} to obtain the Higgs boson mass and the 
electroweak precision observables. We apply \textsf{GM2Calc} \cite{Athron:2015rva} for the calculation 
of the muon $g-2$. Finally, we interface all these programs with \textsf{EasyScan\_HEP} \cite{es} and 
adopt the built-in Metropolis-Hastings algorithm to perform the MCMC scan.

\begin{table}[t] \centering\caption{The experimental constraints used in the likelihood function $\mathcal{L}_{\textrm{precision}}$. \label{tab:experiment}}
\begin{tabular}{|c|c|c|c|c|c|c|}
\hline  Observable					                     & $\mu$		& $\sigma$(exp)	& $\tau$(th)	& References \\
\hline  $\Omega_{\chi}h^2$				                 & 0.1186		& 0.0031		& 0.012		& \cite{Ade:2013zuv} \\
\hline  $\Delta a_{\mu}\times 10^{10}$ 			     & 26.1		    & 8		    & 2		&\cite{Hagiwara:2011af}   \\
\hline  $BR(B\rightarrow X_s\gamma)\times10^4$		     & 3.43		& 0.22		& 0.24		&   \cite{Amhis:2014hma}  \\
\hline  $BR(B_s^0 \rightarrow \mu^+\mu^-)\times10^9$	 & 2.9		& 0.7		& 0.29		&  \cite{CMS:2014xfa} \\
\hline  $R(B^-\rightarrow \tau^- \hat{\nu}_{\tau})$& 1.04	& 0.24		& 0.24 		& \cite{Amhis:2014hma}  \\
\hline  $m_h$ (GeV)						                    & 125.36		& 0.41 		& 2.0		&  \cite{Aad:2014aba} \\
\hline  $M_W$	(GeV)					                    & 80.385		& 0.015		& 0.01		&  \cite{PDG} \\
\hline  $sin^2\hat{\theta}(M_Z)(\overline{MS})$		 & 0.23153	& 0.00016	& 0.00010	& \cite{ALEPH:2005ab} \\
\hline
\end{tabular}
\end{table}

Since the LHC searches for the jets and missing transverse momentum, with or without leptons or $b$-jets 
events, give strong constraints on the CMSSM parameter space \cite{Aad:2015iea},  we implement these 
bounds by recasting the relevant experimental analyses. We simulate the SUSY signal processes with 
\textsf{MG5@NLO} \cite{Alwall:2014hca}, PYTHIA \cite{Sjostrand:2006za} and Delphes3.3.0 \cite{Ovyn:2009tx}. 
We use \textsf{Prospino} \cite{Beenakker:1996ed} to calculate the cross sections of the squark and gluino 
productions at the next-to-leading order.
The likelihood function $\mathcal{L}_{\textrm{LHC}}$ is evaluated by assuming a Poisson distribution of the 
expected signal events $N_s$, the observed events $N_o$ and the background events $N_b$ from the 
experimental reports.

\begin{figure}[]
  \centering
  \includegraphics[height=5cm]{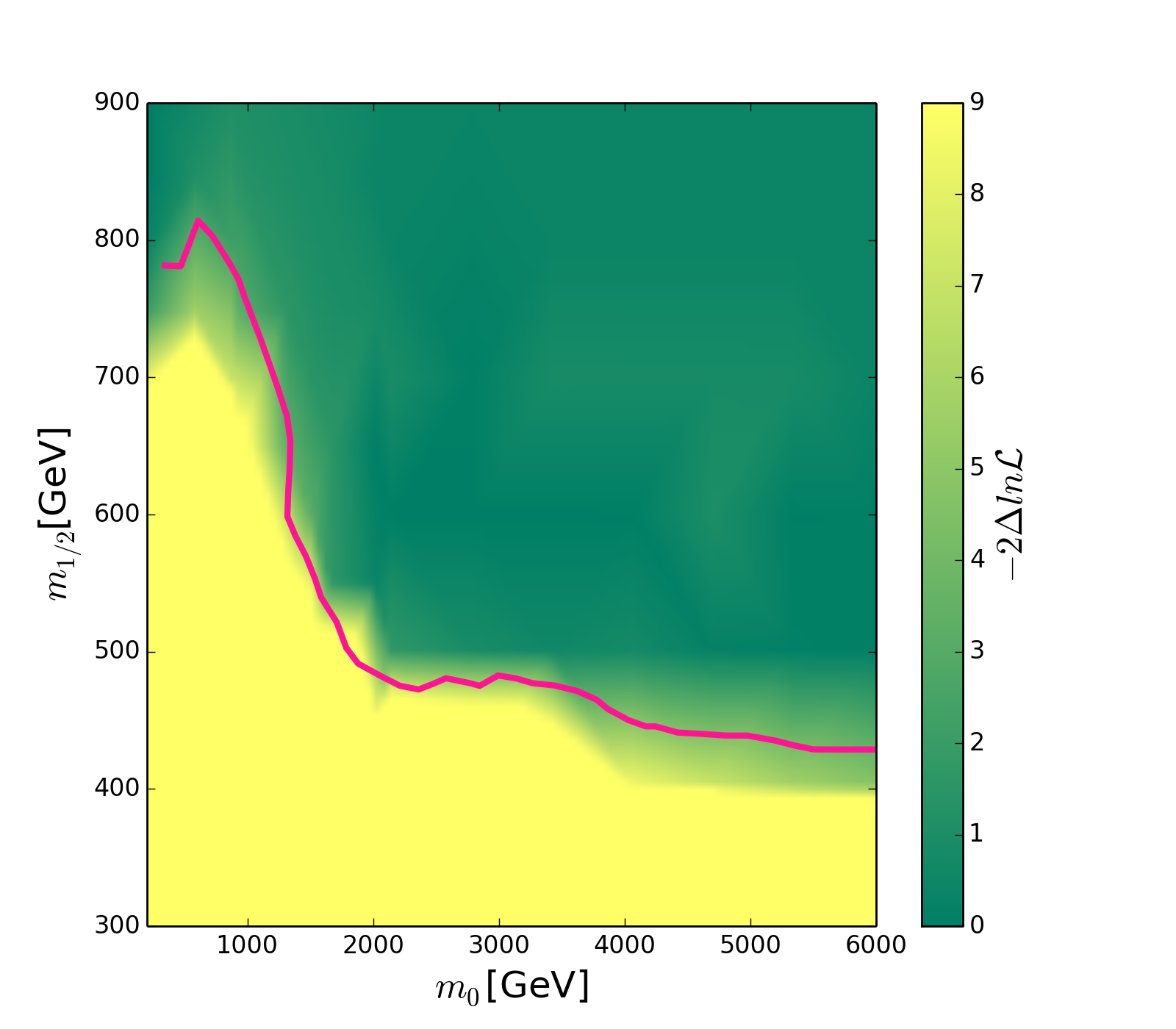}
  \includegraphics[height=5cm]{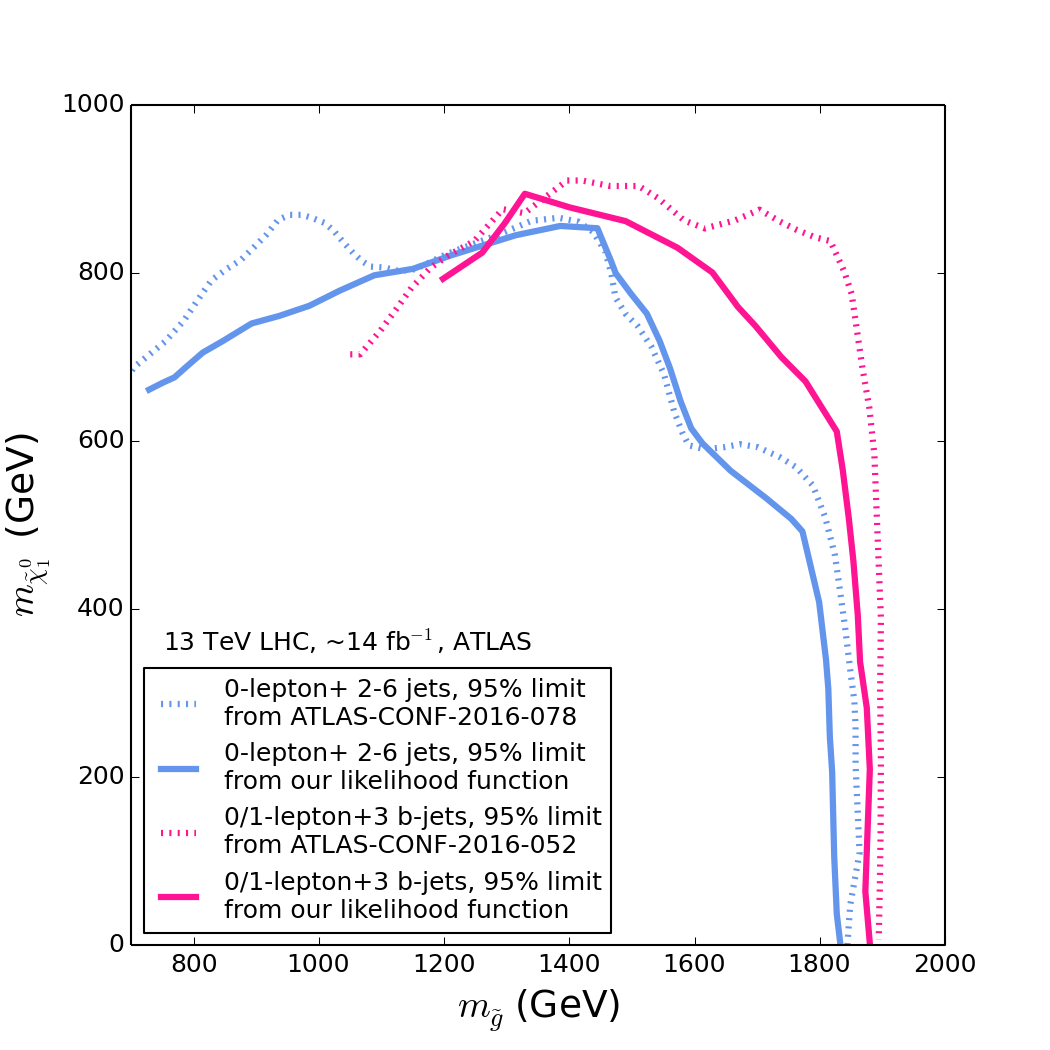}
  \includegraphics[height=5cm]{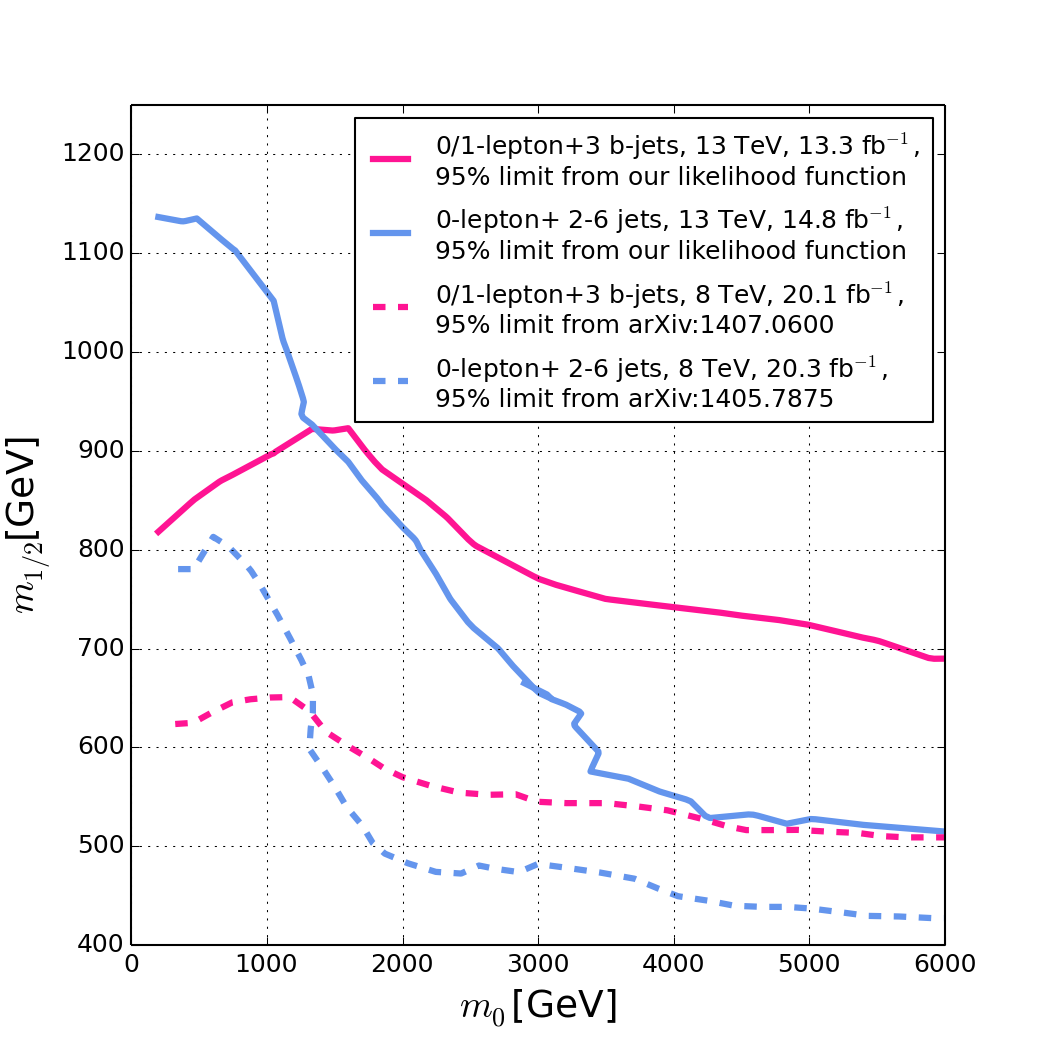}
\vspace{-0.5cm}
  \caption{The comparisons of the ATLAS exclusion limits with our simulation results at the 8 
and 13 TeV LHC. In the left panel, the curve is the ATLAS  95\% CL limits from the 
0 lepton + 2-6 jets +$E_T^{\rm miss}$ search while the  contour map is our  simulation results
( $-2\ln\mathcal{L} =4$ corresponds to the 95\% limits).     
}\label{fig:atlas}
\end{figure}
To validate our method, we firstly calculate the likelihood $\mathcal{L}_{\textrm{8 TeV LHC}}$ by recasting the 
ATLAS analyses 0 lepton + 2-6 jets +$E_T^{\rm miss}$ \cite{Aad:2014wea} and 
0/1 lepton + 3 b-jets + $E_T^{\rm miss}$ \cite{Aad:2014lra} with CheckMATE 2.0 \cite{Drees:2013wra} on 
the $(m_0,m_{1/2})$ plane of the CMSSM. Other parameters are fixed as $\tan\beta = 30$, 
$A_0 = -2m_0$, $\mu> 0$. In the left panel of Fig.\ref{fig:atlas}, we show the color map 
of $-2\Delta\ln\mathcal{L}$ on $(M_0,M_{1/2})$ plane for 0 lepton + 2-6 jets +$E_T^{\rm miss}$ search. 
We can see that the contour of $-2\ln\mathcal{L} =4$ (our 59\% limits) is well consistent with the 
experimental 95\% CL limits. 
In the middle panel of Fig.\ref{fig:atlas}, we also validate our simulations for the ATLAS 13 TeV results 
of 0 lepton + 2-6 jets +$E_T^{\rm miss}$ \cite{ATLAS:2016kts} 
and 0/1 lepton + 3 $b$-jets + $E_T^{\rm miss}$ \cite{ATLAS:2016uzr} on the plane of $m_{\tilde{g}}$ 
versus $m_{\tilde{\chi}^0_1}$. It can be seen that our 95\% CL exclusion limits for the simplified models 
are well consistent with the ATLAS results.  Since the signal rate is fairly insensitive to tan$\beta$ 
and $A_0$ \cite{Allanach:2011ut}, we can only take the likelihood as a function of  $(m_0,m_{1/2})$. Then, 
the likelihood function $\mathcal{L}_{\textrm{13 TeV LHC}}$ is built by interpolating the likelihood of the grid 
points on $(m_0,m_{1/2})$ plane with interval of 300 GeV and 50 GeV for $m_0$ and $m_{1/2}$, respectively. 
In the right panel of Fig.\ref{fig:atlas}, we compare our 95\% C.L. exclusion limits from the 13 TeV 
analyses with those from the 8 TeV analyses. We can find that the lower values of $M_{1/2}$ is lifted 
from 800 GeV at the 8 TeV LHC to 1100 GeV at the 13 TeV LHC.

The likelihood function $\mathcal{L}_{\textrm{LEP}}$ from the LEP direct searches for sparticles is evaluated 
by using a step function. We set a large value of $\ln\mathcal{L}_{\textrm{LEP}}$ for the samples excluded by 
LEP, while set 0 for the samples allowed. We find that our results are not sensitive to 
$\mathcal{L}_{\textrm{LEP}}$ because LEP limits on the sparticle masses are much weaker than those from the LHC.

The likelihood function $\mathcal{L}_{\textrm{LUX-2013}}$ arising from the results of the LUX with 85.3 live 
days of data collected between April and August 2013 with a 118 kg fiducial volume is computed by 
the \textsf{LUXCalc} package \cite{Savage:2015xta}. For the recent direct detection constraints from 
LUX using a 3.35e4 kg-day exposure,  we use the form of likelihood function described 
in \cite{deAustri:2006jwj} to estimate $\mathcal{L}_{\textrm{LUX 2016}}$ with a theoretical uncertainty 
of 10\%. This form of likelihood function is built for the experimental result presented in only 
upper or lower bounds. The spin-independent dark matter-nucleon scattering cross section used in the 
likelihood function is obtained from \textsf{micrOMEGAs} with $f_{Tu} = 0.023689$, $f_{Td} = 0.03906$ 
and $f_{Tg} = 0.363$.

To show the impact of the recent LHC 13 TeV direct searches for sparticles and the LUX dark matter 
detections on the parameter space of the CMSSM,
we use two likelihood functions:
\begin{equation}
\begin{aligned}
    &\ln\mathcal{L}(\eta)^{\textrm{old}}=\ln\mathcal{L}_{\textrm{precision}}+\ln\mathcal{L}_{\textrm{8 TeV LHC}}+\ln\mathcal{L}_{\textrm{LEP}}+\ln\mathcal{L}_{\textrm{LUX 2013}},\\
    &\ln\mathcal{L}(\eta)^{\textrm{new}}=\ln\mathcal{L}(\eta)^{\textrm{old}}+\ln\mathcal{L}_{\textrm{13 TeV LHC}}+\ln\mathcal{L}_{\textrm{LUX 2016}},
\end{aligned}
\end{equation}

\subsection{Profile likelihood function}

There are two statistical ways to show the global fit results:
the marginal posterior PDF and the profile likelihood function.
The marginal posterior PDF of a given parameter $\eta_i$ is defined by integrating out the posterior
distribution $p(\eta|d)$  of other parameters in Eq.(\ref{eq:bayes}):
\begin{equation}
\begin{aligned} \label{stat:Bayes}
   & p(\eta_i|d)=\int{p(\eta|d)d\eta_1 ... d\eta_{i-1} d\eta_{i+1} ... d\eta_n},\\
\end{aligned}
\end{equation}
which includes the volume effects and the peaks of the highest posterior mass. However, the result from 
the marginal posterior PDF depends on the assumption of different prior probabilities.

The frequentist profile likelihood function is defined as
\begin{equation}
\begin{aligned} \label{stat:Frequen}
   & \mathcal{L}(\eta_i)=\mathop{\max}_{\xi}\mathcal{L}(\eta)=\mathcal{L}(\eta,\hat{\hat{\xi}} ),\\
\end{aligned}
\end{equation}
where $\mathcal{L}(\eta)$ is the likelihood function in Eq.(\ref{eq:bayes}),
$\xi = \eta_1, ..., \eta_{i-1}, \eta_{i+1}, ..., \eta_n$ and $\hat{\hat{\xi}}$ is the value of $\xi$
which maximizes $\mathcal{L}(\eta)$. According to the definition, the profile likelihood
function is independent of the scan method. Thus the samples obtained from MCMC scans with `\textit{Flat}'
or `\textit{Log}' prior PDF can be merged together. Since the frequentist approach leads to a well defined 
probability, we present our results by the profile likelihood. The confidence intervals/regions from the 
resulting 1D/2D profile likelihood maps are given by the profile likelihood ratio 
$-2\ln\lambda(\eta_i) = -2\ln(\mathcal{L}(\eta_i)/\mathcal{L}(\hat{\eta_i},\hat{\xi}))$ which
is $\chi^2$-distributed, with $\mathcal{L}(\hat{\eta_i},\hat{\xi})$ corresponding to the maximum 
likelihood function (i.e. the best fit point).

\section{Results}
In this section, we present the allowed parameter space of the CMSSM from the profile likelihood functions 
of 1 billion samples.

\begin{figure}[]
  \centering
  \includegraphics[width=8cm]{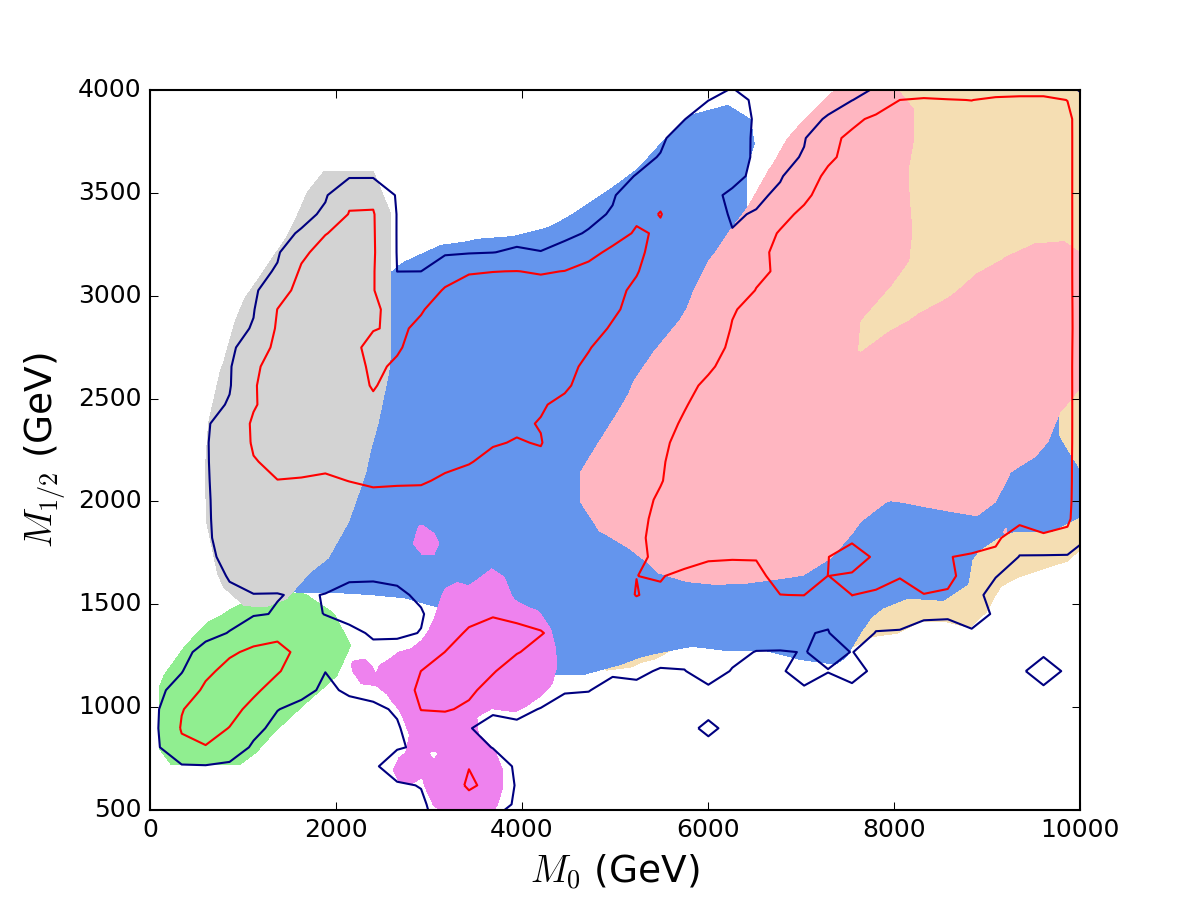}  \includegraphics[width=8cm]{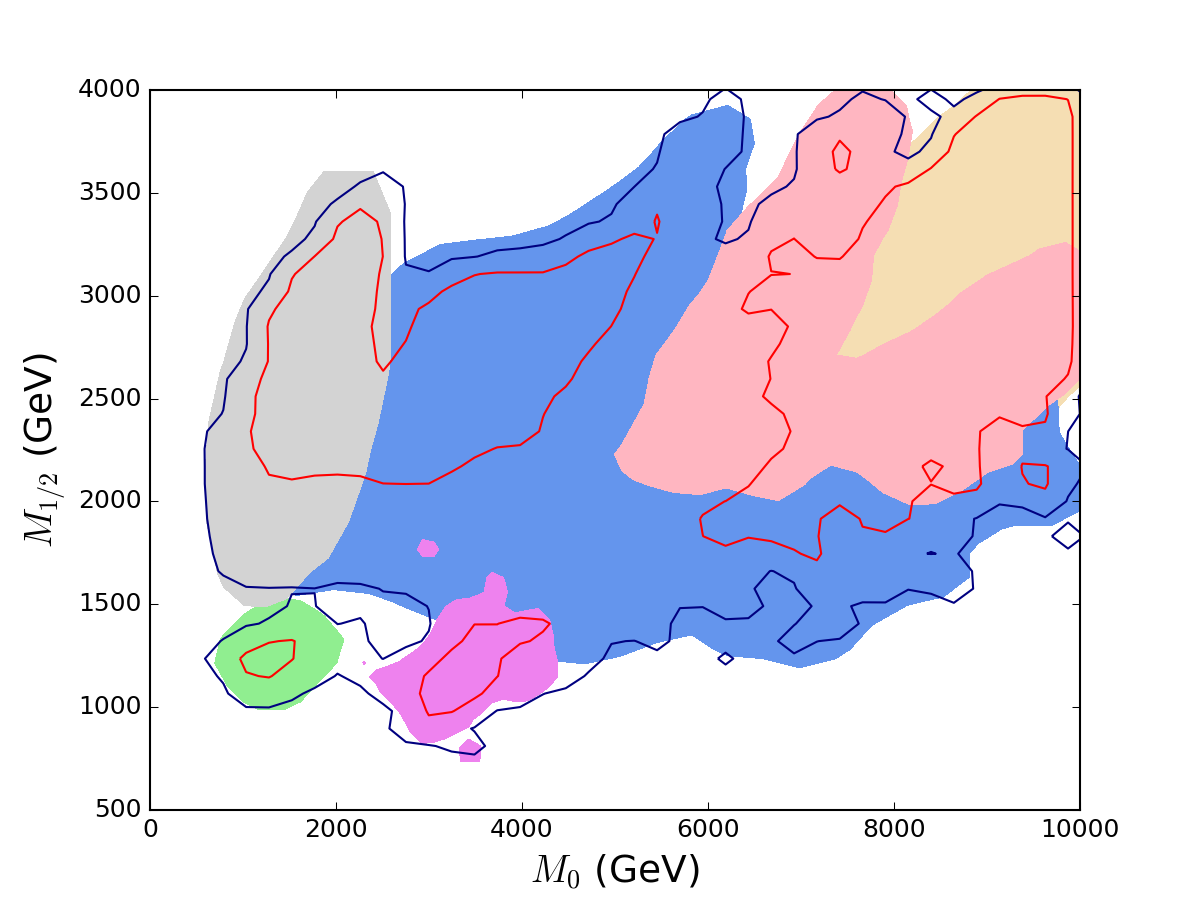}\\
  \includegraphics[width=8cm]{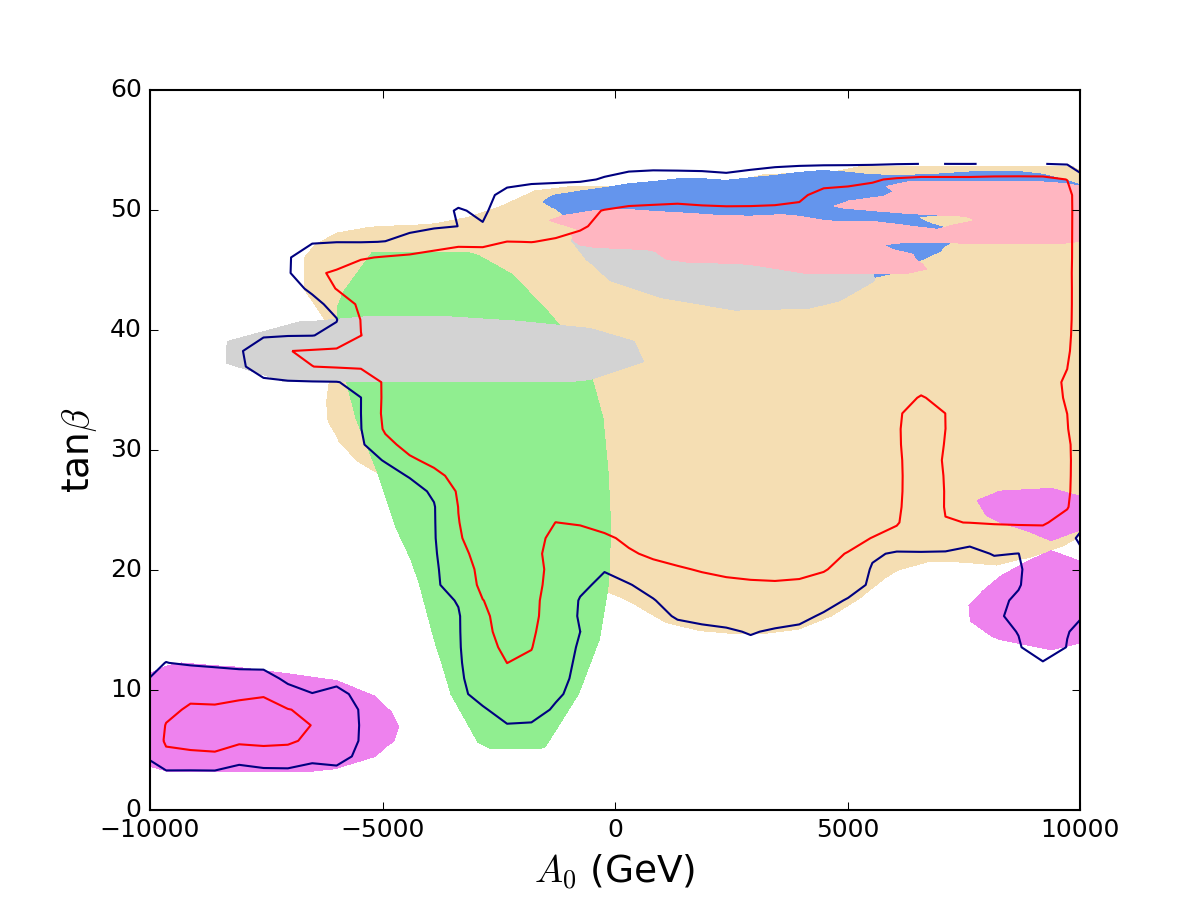}  \includegraphics[width=8cm]{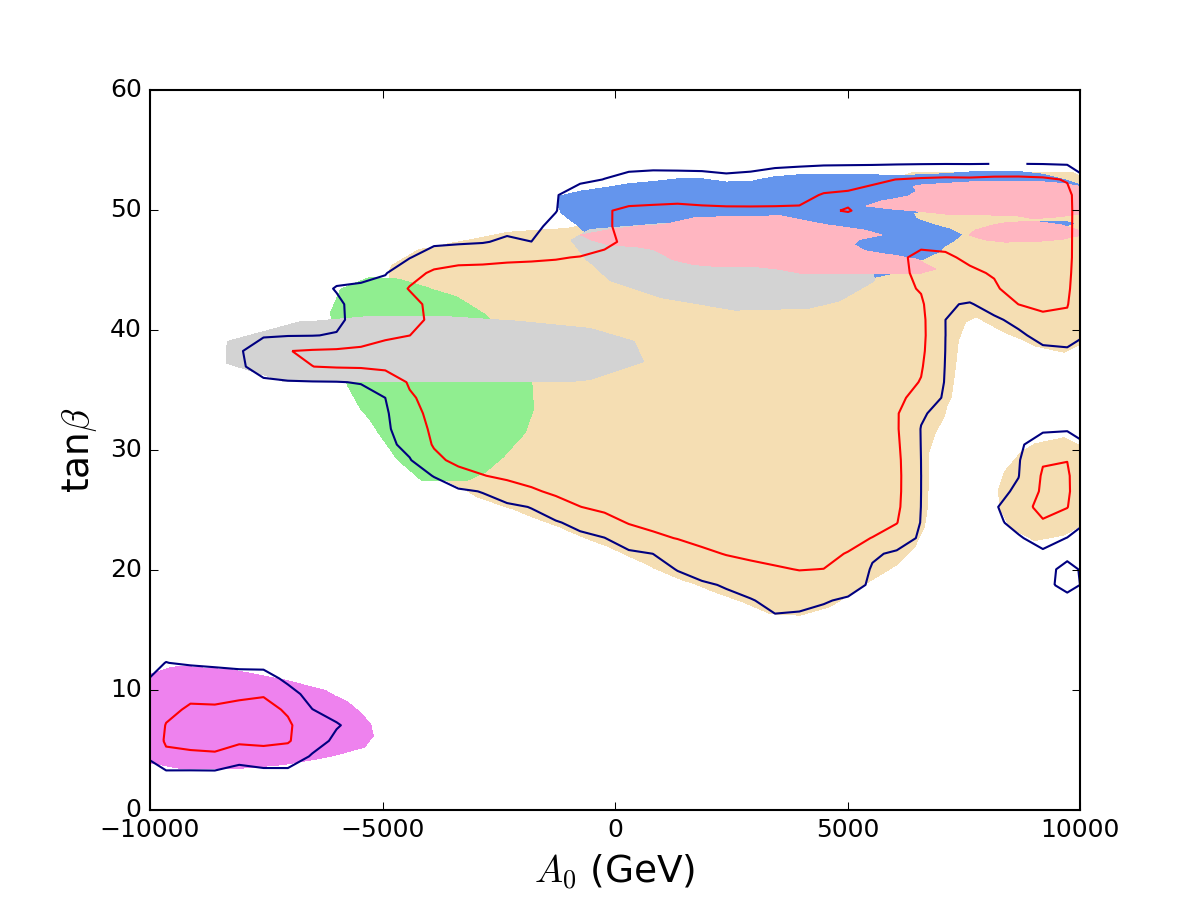}\\
  \includegraphics[width=8cm]{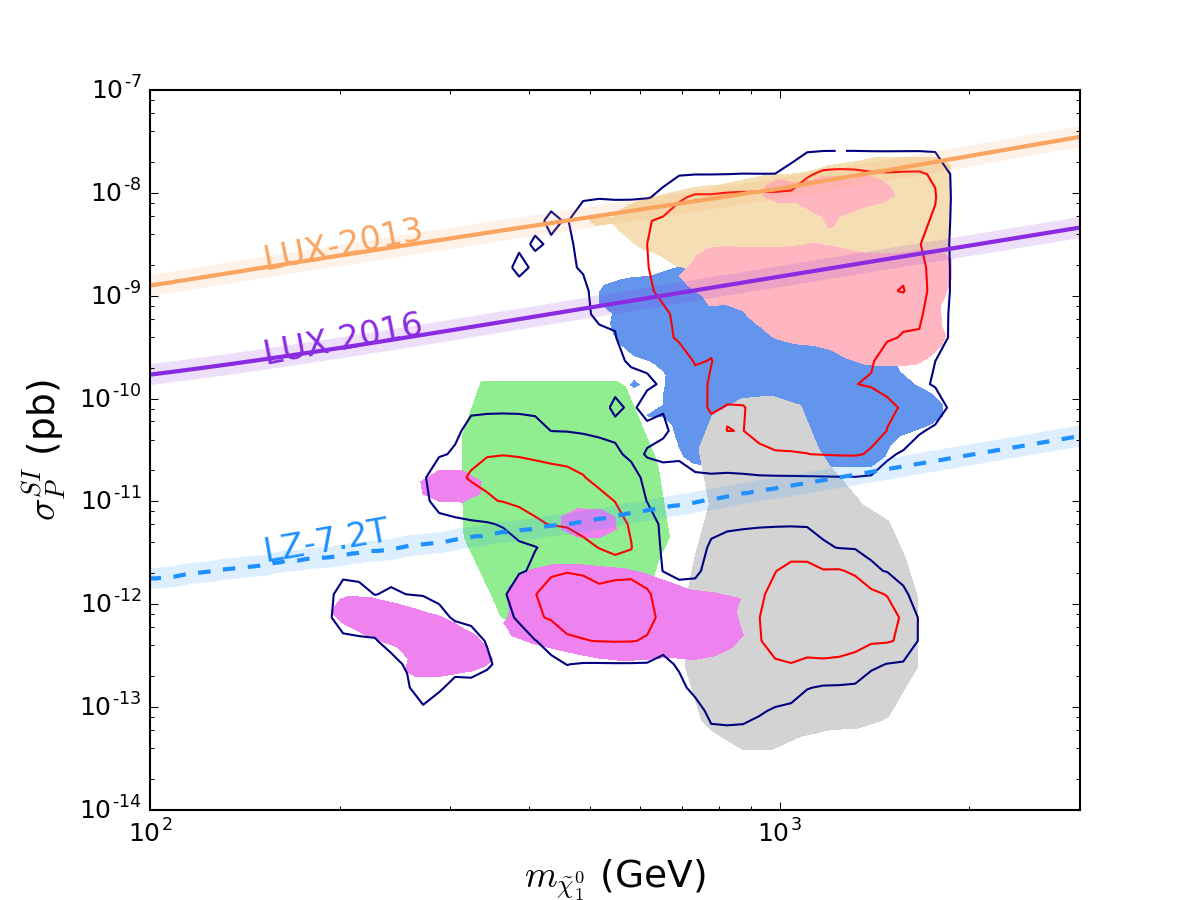}  \includegraphics[width=8cm]{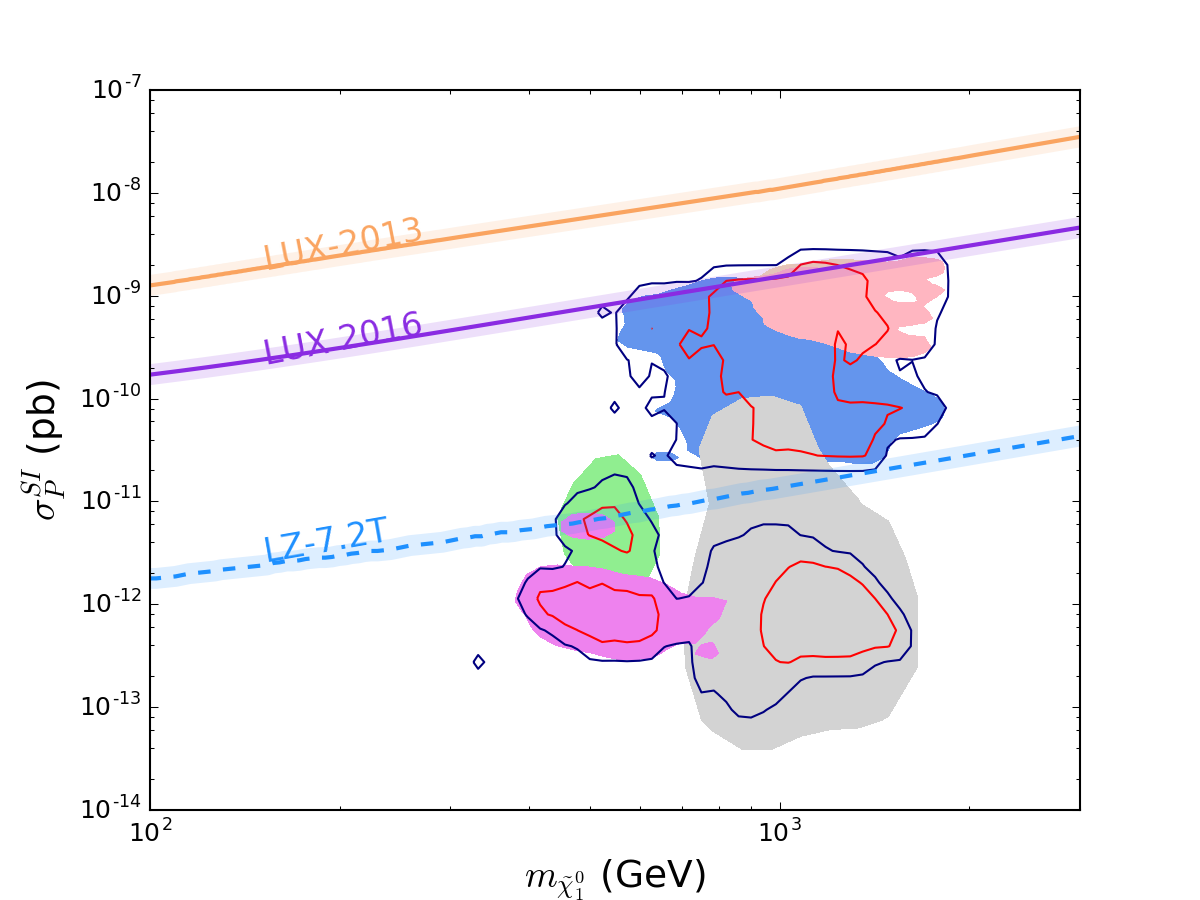}\\
  \caption{
The allowed parameter regions of the CMSSM. 
The red and blue contours correspond to 68\% and 95\% CL regions obtained 
from $\mathcal{L}(\eta_{i})^{\textrm{old}}$ (left column) and $\mathcal{L}(\eta_{i})^{\textrm{new}}$ (right column). 
The filled colors correspond to the main six dark matter annihilation mechanisms: 
StauC (green), {StopC (purple),} AF (blue), FP (yellow), StauC$\&$AF (gray) and AF$\&$FP (pink). 
The 90\% exclusion limits on the spin-independent dark matter-nucleon scattering cross-section 
($\sigma^{\rm SI}_p$) from LUX 2013 \cite{Akerib:2013tjd}, LUX 2016 \cite{Akerib:2016vxi} and LZ \cite{LZ} 
are also displayed in the bottom panels.
   }\label{fig:2D}
\end{figure}

In Fig.~\ref{fig:2D}, we show the allowed parameter region of CMSSM on the planes of 
$(m_{1/2},m_0)$, $(\tan\beta,A_0)$, and $(m_{\tilde{\chi}_1^0},\sigma_p^{\rm SI})$, respectively. 
The red and black contours correspond to 68\% and 95\% CL regions. The left and right columns depict 
the 2D profile likelihood functions obtained by $\mathcal{L}(\eta)^{\textrm{old}}$ and 
$\mathcal{L}(\eta)^{\textrm{new}}$, respectively. {Six} regions that have different DM annihilation mechanisms 
are distinguished by colors \cite{Bagnaschi:2015eha}: the StauC region (green), in which the relic density 
of the neutralino is derived by the neutalino-stau co-annihilation with 
$m_{\tilde{\tau}_1}/m_{\tilde{\chi}_1^0} -1<0.15$; 
{ the StopC region (purple), where one of the stops is the next-to-lightest sparticle and slightly heavier than the LSP neutralino($m_{\tilde{t}_1}/m_{\tilde{\chi}_1^0} -1<0.2$), so that they can co-annihilate to get the right relic density \cite{Boehm:1999bj,Ellis:2001nx,Santoso:2002xu,Edsjo:2003us,Chen:2010kq,Ajaib:2011hs,Yu:2012kj,Harz:2012fz,Ellis:2014ipa,Harz:2014tma,Raza:2014upa,Buchmueller:2015uqa,Bagnaschi:2015eha,Kaufman:2015nda}; }
the AF region (blue) is characterized by 
$|m_{A}/2m_{\tilde{\chi}_1^0} -1|<0.2$, where a relatively light pseudoscalar can be the mediator for 
DM annihilation in $s$-channel; the FP region (yellow) requires $|\mu/m_{\tilde{\chi}_1^0} -1|<0.4$,  
where the lightest neutralino has a significant higgsino component. 
Besides, two hybrid regions, StauC$\&$AF (gray) and AF$\&$FP (pink), which satisfy two of the above conditions 
simultaneously.

From Fig.~\ref{fig:2D}, we find that the parameter space of these six regions is sensitive to different 
experiments:
\begin{itemize}
  \item In the StauC region is characterized of $M_0<2$ TeV and $M_{1/2}<$ 1.5 TeV with a negative $A_0$ {and positive $\mu$}, 
which corresponds to $m_{\tilde{g}}<3$ TeV and $m_{\tilde{t}_1}<1.8$ TeV. 
{The parameter space of negative $\mu$ is restricted by the discrepancy between the experimental value of $a_{\mu}$ and the SM calculation, and also unfavored by the measurements of BR$(b\to s\gamma)$ and $m_h$.}
From Fig.~\ref{fig:2D}, we find most of {the StauC region} is 
excluded by the recent LHC-13 TeV SUSY direct searches. The small survived region can be further probed at 
the high luminosity LHC. It should be noted that the best point, which previously  
located in the StauC region \cite{Bagnaschi:2015eha,Bechtle:2015nua}, has been pushed into the FP region 
by the new LHC-13 TeV results.
 Besides, the 95\% CL lower limits of $M_0$ and $M_{1/2}$ become 1041 GeV and 856 GeV, respectively.  
As a consequence, the lower masses of $\tilde{g}$ and $\tilde{t}_1$ are pushed up to 2.2 TeV and 1 TeV, 
respectively. The value of $\tan\beta$ is required to be larger than 30 to enhance the stau co-annihilation 
cross section. In addition, when $\tilde{\chi}^0_1$ is bino-dominated, LUX constraints on this region 
will become weak. 
It should also be noted that in this region the mass difference $\delta m_{\tilde{\tau}\tilde{\chi}}$ 
between $\tilde{\tau}_1$ and $\tilde{\chi}^0_1$ can be very small.  As a result, $\tilde{\tau}_1$ may decay 
in the outer part of the LHC detector or outside the detector. It is shown in \cite{Citron:2012fg} that 
the $\tilde{\tau}_1$ signature at the LHC would be a massive metastable charged particle if 
$\delta m_{\tilde{\tau}\tilde{\chi}}<1.2$ GeV and a disappearing track if $\delta m_{\tilde{\tau}\tilde{\chi}}>1.2$ GeV. 
The exclusion limits yielded from the searches for disappearing tracks and metastable charged particles at the LHC 
have been studied in \cite{Desai:2014uha}. As shown in \cite{Desai:2014uha}, such limits are weaker than those 
from the 0 lepton + 2-6 jets +$E_T^{\rm miss}$ search. For example, the CMS direct searches for long-lived charged 
particles set a lower limit of 340 GeV on a stable stau with the LHC Run-1 data \cite{Chatrchyan:2013oca} and 
about 400 GeV with 12.9 fb$^{-1}$ data at the 13 TeV LHC \cite{CMS:2016ybj}, while in our global fit the 95\% CL 
lower limits on stau mass are 351 GeV and 475 GeV obtained from $\ln\mathcal{L}(\eta)^{\textrm{old}}$ and 
$\ln\mathcal{L}(\eta)^{\textrm{new}}$, respectively.

  \item Contrary to the StauC region, the FP region prefers large values of $M_0$ and $M_{1/2}$ {with both positive and negative $\mu$}.  
So such a region can escape the LHC SUSY search limits. However, in this region $\tilde{\chi}^0_1$ 
is higgsino-dominated and the spin-independent $\tilde{\chi}^0_1$-nucleon scattering cross-section 
is sizable. We can see that 1/3 of such a parameter space is excluded by LUX-2016. The whole region 
lies in the expected exclusion limits of future LZ experiment. 

 \item {In the StopC region, the mass difference between the stop and LSP is restricted to a very narrow strip,  $m_{\tilde{t}_1}-m_{\tilde{\chi}_1^0} <  30 \sim 44$ GeV where $m_{\tilde{\chi}_1^0}$ is between $300\sim900$ GeV, which makes this region hard to be fully explored by LHC search. In this region, the LSP neutralino is almost exclusively a bino with mass $m_{\tilde{\chi}_1^0}\simeq 0.4 M_{1/2} \sim (300\sim900)$ GeV, where the lower bound comes from LHC SUSY direct searches and the upper bound is derived from the limited range of $A_0$. A rather large $A_0$, which appears in the off-diagonal term of stop square mass matrix, is required to get a rather light $\tilde{t}_1$.  Since the wino and Higgsino component of LSP neutralino is extremely tiny, proton-neutralino scattering cross sections are expected to be small. Therefore, the DM direct scarcely restrict the StopC region. }

  \item In the AF region, the $\tilde{\chi}^0_1$ pair mostly annihilate into $b$-quarks through 
$A^0$ boson in the $s$-channel. So a large value of $\tan\beta$ is preferred because it can enhance the 
coupling between $A^0$ and $b$ quarks. The preferred parameter space by  
$\mathcal{L}(\eta)^{\textrm{old}}$ and $\mathcal{L}(\eta)^{\textrm{new}}$ are scarcely changed. 
Another special feature of the AF region is that the profile likelihood is not as flat as the SC and 
FP regions on the ($M_0$, $M_{1/2}$) plane. We can see that the 95\% CL and 68\% CL contours are very 
close in the SC and FP regions, while the 95\% CL region is much lager than the 68\% CL region in 
the AF region. This is mainly because that the observable $\mathop{\rm Br}(B_s\to\mu^+\mu^-)$ is proportional 
to $\tan^6\!\beta$. The whole AF region can also be covered by the LZ experiment.
In addition,  the study in \cite{Kowalska:2013hha} showed 
 that the AF region would be covered by the $\mathop{\rm Br}(B_s\to\mu^+\mu^-)$ measurement at the 
14 TeV LHC with 50 fb$^{-1}$ luminosity data.
  \item { Besides above pure regions, there are two `hybrid' regions satisfy two of the conditions. 
  However, they basically dominated by signal annihilation mechanism. The pink region have similar properties to the AF region, i.e., 
 moderate $M_0$, $M_{1/2}$, a large tan$\beta$ and a positive $A_0$. The samples in the gray region generate appropriate DM relic density mainly by stau co-annihilation. But they are much different from the StauC region as their signs of $\mu$ are negative. As a result, they are blind to both LHC and DM direct detection experiments. Fortunately, negative $\mu$ leads to worser $\Delta a_{\mu}$ than the SM prediction, which retains the possibility to detect this region. }
\end{itemize}

\begin{figure}[]
  \centering
  \includegraphics[height=5.1cm]{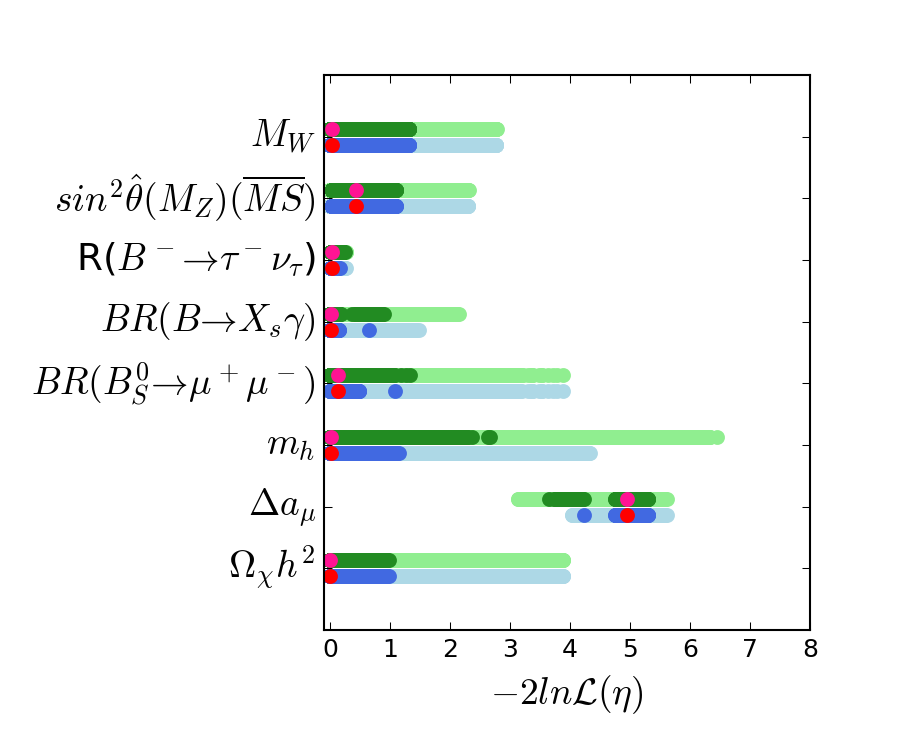} \includegraphics[height=5.1cm]{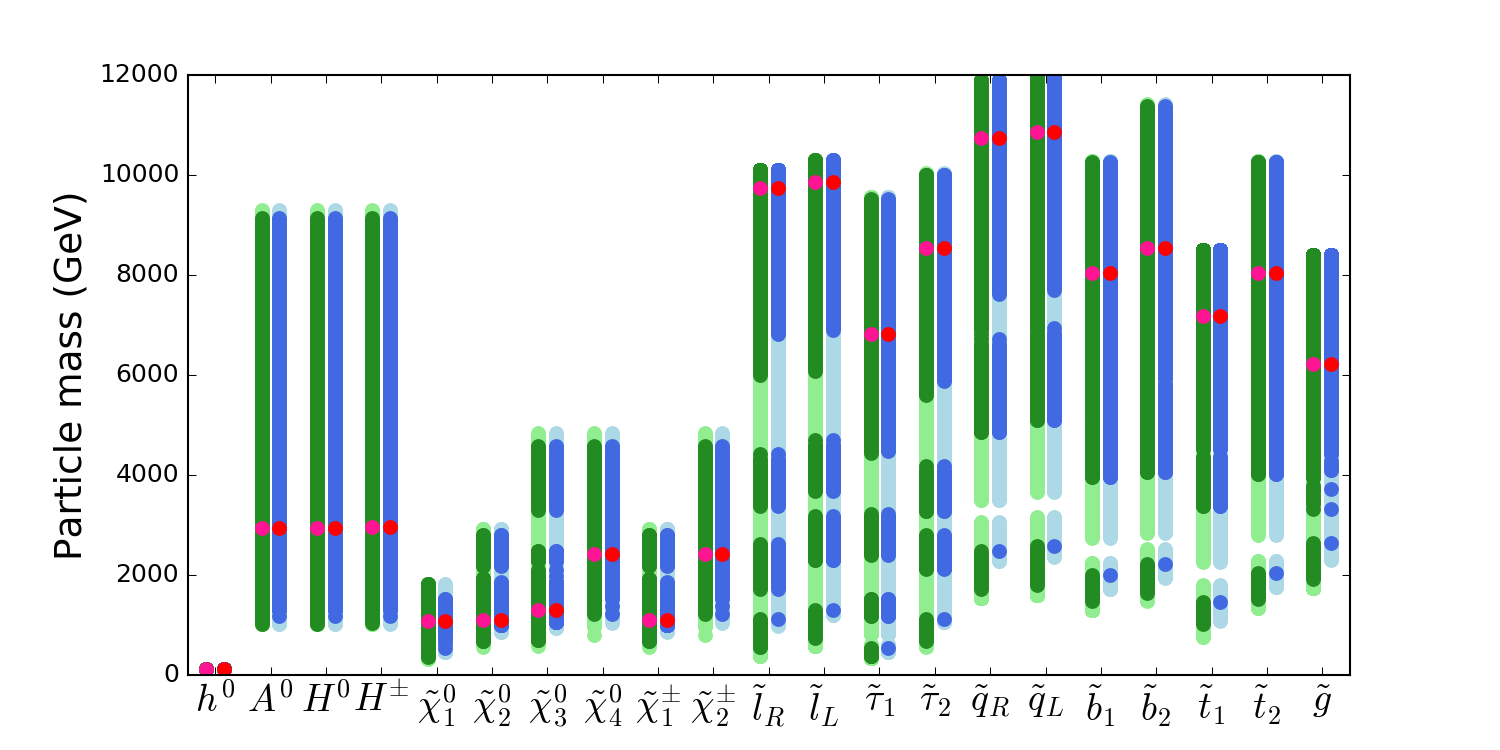}
\vspace*{-.8cm} 
 \caption{The 68\% and 95\% CL intervals indicated by light and dark green (blue) bars for the 1D 
profile likelihood functions obtained from 
$\mathcal{L}(\eta_{i})^{\textrm{old}}$ ($\mathcal{L}(\eta_{i})^{\textrm{new}}$), where $i$ stands the low energy 
observations in left panel and  masses of SUSY particles in the right panel. The red points represent 
the values of the over-all best-fit point.}\label{fig:1D}
\end{figure}

In Fig.~\ref{fig:1D}, we display the predictions of the low energy observables (left panel) and mass 
spectrum of sparticles (right panel) in the preferred parameter space.  The light and dark green (blue) 
bars indicate 95\% and 68\% CL 1D intervals obtained from $\mathcal{L}(\eta)^{\textrm{old}}$ 
($\mathcal{L}(\eta)^{\textrm{new}}$), while the red points stand for the over-all best-fit point. 
From Fig.\ref{fig:1D} we can see that the main contribution to the likelihood comes from the muon 
$g-2$, $\delta a_{\mu}^{\rm SUSY}$. Due to the new LHC strong constraints on the StauC region, the SUSY contribution 
to $\Delta a_{\mu}$ becomes rather small. The 95\% CL upper limit on $\delta a_{\mu}^{\rm SUSY}$ is reduced 
from $6\times10^{-10}$ ($\mathcal{L}(\eta)^{\textrm{old}}$) to $3\times10^{-10}$($\mathcal{L}(\eta)^{\textrm{new}}$). 
{Compared with the LHC Run-1, the null results from the 13 TeV LHC SUSY searches also push the 95\% CL 
lower mass bounds of the gluino and the first two generations of squarks from about 1.2 TeV to 1.8 TeV, 
the lightest stop from 232 GeV to 356 GeV, and the LSP from 196 GeV to 324 GeV.} 
Therefore, the SUSY 
contributions to $B$-physics observables also approach to the decoupling limit. The Higgs mass 
is much easier to satisfy the experimental measurements. Note that, except for electroweakinos, the upper 
mass limits on sparticles are determined by the scan range of the input parameters, i.e., the prior PDF.

\begin{table}[]
\centering
\caption{The properties of the benchmark points chosen from the StauC, FP, AF, StopC and Hybrid regions with 
local $\mathcal{L}(\eta)^{\textrm{new}}_{\textrm{local}}$, where StauC$'$ represent the StauC region 
with local $\mathcal{L}(\eta)^{\textrm{old}}_{\textrm{local}}$. Each benchmark point is the best point in
the corresponding region. 
$\Delta$ is the fine-tuning extent. 
{$(\mu)$ stands for sign$(\mu)$.}
All masses are in unit of GeV.
}
\label{tab:bf}
\begin{tabular}{|l|c|c|c|c|c|c|c|c|c|c|c|c|c|c|c|c|c|c|c|c|c|c|}
\hline
Point     & $M_0$ & $M_{1/2}$  & $A_0$    & tan$\beta$ & $(\mu)$ &$m_h$   & $m_A$  & $m_{\tilde{\chi}_1^0}$  & $m_{\tilde{\tau}_1}$ &  $m_{\tilde{\chi}_1^{\pm}}$ &  $\Omega h^2$   & $\sigma_P^{SI}$    & $\Delta a_{\mu}$     & $\mathcal{L}(\eta)$ &  $\Delta$   \\ \hline
~FP  & 9701  & 2881 & 8869  & 50.3 & $+$ & 125.1 & 2947 & 1089 & 6823  & 1095 & 0.119 & 1.4E-09 & 2.3E-11 & 10.6    & 4315 \\ \hline
~AF  & 8925  & 2598 & 9531  & 51.2 & $+$ & 124.6 & 2488 & 1167 & 5947  & 1691 & 0.125 & 1.9E-10 & 2.8E-11 & 10.9    & 3445 \\ \hline
~StopC &  4145 & 1400 & -9891  & 5.6  & $-$  & 125.3 & 6116 & 628   & 4139  & 1199 & 0.122 & 7.77.E-13 & -3.0E-12 & 11.0    & 11817   \\ \hline
~StauC & 1026  & 1221 & -3397 & 31.5 & $+$ & 123.6 & 1928 & 531  & 531   & 1005 & 0.123 & 5.9E-12 & 2.3E-10 & 11.5    & 2106   \\ \hline
~StauC$'$ & 728   & 1110 & -2774 & 26.3 & $+$ & 123.1 & 1800 & 479  & 479   & 908  & 0.120 & 8.6E-12 & 2.9E-10 & 11.1    & 1641 \\ \hline
~Hybrid &  1872 & 3199 & -4897  & 37.8 & $-$ & 125.4 & 2923 & 1440 & 1441  & 2662 & 0.118 & 8.81E-13 & -5.2E-11 & 11.1    & 8079 \\ \hline
\end{tabular}
\end{table}
In Table \ref{tab:bf}, we present the properties of the benchmark points chosen from the StauC, StopC, FP, AF and Hybrid regions
 with local $\mathcal{L}(\eta)^{\textrm{new}}_{\textrm{local}}$, where StauC$'$ stands for the StauC region with local 
$\mathcal{L}(\eta)^{\textrm{old}}_{\textrm{local}}$. We can see that the local best point in the StauC region 
has a smaller Higgs mass but a larger muon $g-2$ than those in FP and AF regions. On the other hand, 
the spin-independent cross section of the local best point in FP region is much larger. 
We also computed the fine-tuning by SOFTSUSY 3.3.1 with the definition \cite{bg}:
\begin{equation}
\begin{aligned}
   \Delta = \mathop{\textrm{max}}_a \left( \left|\frac{\partial \ln M^2_Z}{\partial \ln a}\right| \right),\\
\end{aligned}
\end{equation}
where $a$ stands for the set of fundamental parameters $\{ M_0, M_{1/2}, A_0, \mu, B \}$ at the GUT scale. 
It can be seen that the fine-tuning value of the local best point in the StauC region is about two thousand, 
which is much smaller than those in FP and AF regions.

\section{Conclusion}
In this paper, we examined the status of the CMSSM by performing a global fit under the current available 
constraints, in particular the LHC-13 TeV SUSY direct searches and the combined LUX Run-3 and -4 dark matter 
direct detection limits.  From the profile likelihood functions of 1 billion samples, we obtained the 
following results for the parameter space of the CMSSM at 95\% confidence level:
(i)  The stau coannihilation(StauC) region has been mostly excluded by the latest LHC Run-2 data; 
(ii) The focus point(FP) region has been largely covered by the LUX-2016 limits while the $A$-funnel(AF) region  
has been severely restricted by flavor observables like $B_s \to \mu^+\mu^-$. The remaining parts of both 
regions will be totally covered by the future LZ dark matter experiment; 
{(iii) Part of the stop coannihilation(StopC) region may be detected with higher integrated luminosity LHC;
(iv) The StauC$\&$AF hybrid region still survives considering both LHC and dark matter experiments;}
{(v) The masses of the stop, the lightest neutralino and the gluino have been pushed up to 363 GeV, 
328 GeV and 1818 GeV, respectively.}
\section*{Acknowledgement}
This work is partly supported by the Australian Research Council, by the National Natural Science
Foundation of China (NNSFC)  under grant Nos. 11675242, by World Premier International Research Center Initiative (WPI Initiative) MEXT of Japan, by the CAS Center for Excellence in Particle Physics (CCEPP)
and by the CAS Key Research Program of Frontier Sciences.

\end{document}